\documentclass[preprint,aps,showpacs,nofootinbib,preprintnumbers,amsmath,amssymb]{revtex4-1}
\usepackage{epsfig}
\usepackage{subfigure}
\usepackage{dcolumn}
\usepackage{bm}
\usepackage[usenames ,dvipsnames]{xcolor}
\usepackage{slashed}
\usepackage{float}
\usepackage{graphicx,color}
\usepackage{amsmath}
\usepackage{nonfloat}
\renewcommand\arraystretch{1}
\begin{document}
\title{Asymmetries of anti-triplet charmed baryon decays }

\author{C.Q. Geng$^{1,2,3}$, Chia-Wei Liu$^{2}$ and Tien-Hsueh Tsai$^{2}$}
\affiliation{
$^{1}$Chongqing University of Posts \& Telecommunications, Chongqing 400065\\
$^{2}$Department of Physics, National Tsing Hua University, Hsinchu 300\\
$^{3}$Physics Division, National Center for Theoretical Sciences, Hsinchu 300
}\date{\today}

\begin{abstract}
We analyze the decay processes of ${\bf B}_c \to {\bf B}_n M$ with the $SU(3)_F$ flavor symmetry and spin-dependent amplitudes, where ${\bf B}_c({\bf B}_n)$ and $M$ are the anti-triplet charmed (octet) baryon and  nonet meson states, respectively. In the $SU(3)_F$  approach, it is the first time that the decay rates and up-down asymmetries are fully and systematically  studied without neglecting the ${\cal O}(\overline{15})$ contributions of the color anti-symmetric parts in the effective Hamiltonian.  Our results of the up-down asymmetries based on $SU(3)_F$  are quite different from the previous theoretical values in the literature.  In particular, we find that the up-down symmetry of $ \alpha(\Lambda_c^+\to \Xi^0 K^+)_{SU(3)} = 0.94^{+0.06}_{-0.11}$, which is consistent with the recent experimental data of $0.77\pm0.78$ by the BESIII Collaboration, but predicted to be zero in the literature.  We  also examine the $K_S^0-K_L^0$ asymmetries  between the decays of ${\bf B}_c \to {\bf B}_n K_S^0$ and ${\bf B}_c \to {\bf B}_n K_L^0$ with  both  Cabibbo-allowed and   doubly  Cabibbo-suppressed transitions.

\end{abstract}
\maketitle

\section{Introduction}
Recently, the Belle collaboration has measured the absolute branching ratio of $\Lambda_c^+ \to pK^-\pi^+$ with high precision~\cite{Belle},
resulting in the world  average value of ${\cal B}(\Lambda_c^+ \to p K^-\pi^+)=(6.23\pm0.33)\%$, given by the Particle Data Group (PDG)~\cite{pdg}. 
 This decay mode  is the so-called golden  channel  as most of  $\Lambda_c^+$ decay branching fractions are presented 
 relative to it. 
 Subsequently, this golden mode, along with  many other $\Lambda_c^+$ ones, has also been observed
by the BESIII Collaboration~\cite{Ablikim:2015flg,Ablikim:2015prg,Ablikim:2016tze,Ablikim:2016mcr,Ablikim:2017ors,
Ablikim:2016vqd,Ablikim:2017iqd,Ablikim:2018jfs,Ablikim:2018bir,Exp:eta'} with   $\Lambda_c^+\bar{\Lambda}_c^-$ pairs, produced by $e^+ e^-$ collisions
at a center-of-mass energy of $\sqrt{s}= 4.6$ GeV, having  a uniquely clean background to study the anti-triplet charmed baryon state of $\Lambda_c^+$. 
 In particular, the decay  of $\Lambda_c^+\to\Sigma^+\eta'$  has been seen for the first time with  $\eta'$ in the final states for the charmed baryon decays~\cite{Exp:eta'}. 
 In addition,  the absolute decay branching fraction of $\Xi_c^0\to \Xi^-\pi^+$, which involves  
 the  anti-triplet charmed baryon state of $\Xi_c^0$, has also been measured  by
 the Belle collaboration~\cite{Exp:absXic0}.  Clearly, a new experimental physics era for charmed baryons has started.
 
 On the other hand,
 the theoretical study of the charmed baryon decays has faced several difficulties. 
 The most serious one is that the  factorization approach in the non-leptonic decays of  charmed baryons 
 is not working. For example, 
 the Cabibbo-allowed decays of $\Lambda_c^+ \to \Sigma^0\pi^+$ and $\Lambda_c^+ \to \Sigma^+\pi^0$ do not receive any factorizable contributions,
 whereas  the experimental  data show that  their branching fractions are all close to $O(10^{-2})$~\cite{pdg}, indicating the failure of the  factorization method.
 In addition, the complication of the charmed baryon structure makes us impossible to directly evaluate the decay amplitude in a model-independent way.
 It is known that  the most reliable and simple way to examine the charmed baryon processes is to use the flavor symmetry of 
$SU(3)_F$~\cite{Sharma:1996sc,Savage:1989qr,Savage:1991wu,Lu:2016ogy,first,zero,second,third,fourth,fifth,Wang:2017gxe,Geng:2019bfz,Zhao:2018mov}.
Indeed, it has been recently demonstrated  that the results for the charmed baryon decays based on the  $SU(3)_F$ 
approach~\cite{Lu:2016ogy,first,zero,second,third,fourth,fifth,Wang:2017gxe,Geng:2019bfz,Zhao:2018mov} are consistent with the current experimental data.
Nevertheless, the charmed baryon decays have been extensively studied  in various dynamical
models~\cite{Korner92,Xu92,Uppal:1994pt,Cheng:general,Cheng:allow,Chen:quark,Cheng:sup,Chen:2002jr,xialpha1,Verma98,xialpha2},
particularly,  the recent dynamical calculations of the singly Cabibbo-suppressed $\Lambda_c^+ $ decays by 
Cheng, Kang and Xu (CKX)~\cite{Cheng:sup}  based on  current algebra.   

For the two-body charmed baryon decay of ${\bf B}_c \to {\bf B}_n M$, 
with ${\bf B}_c({\bf B}_n)$ and $M$  the anti-triplet charmed (octet) baryon and  nonet meson states, respectively,
beside its decay branching fraction, there exits another interesting physical observable, the up-down asymmetry $\alpha$, which is related to
the longitudinal polarization of ${\bf B}_n$. 
Currently, there are three experimental measurements  of the up-down asymmetries in the charmed baryon decays~\cite{pdg},
along with the recent one by BESIII~\cite{Ablikim:2018bir}, given by
\begin{equation}\label{expal}
  \alpha(\Lambda_c^+\to\Xi^0 K^+)_{exp}=0.77\pm0.78\,,
\end{equation}
which has been suggested to be approximately zero in the previous theoretical studies with the 
dynamical models~\cite{Korner92,Xu92,xialpha1,xialpha2,Cheng:allow,Cheng:general,Uppal:1994pt,Verma98}
as well as the $SU(3)_F$ approach~\cite{Sharma:1996sc}.
However, the up-down asymmetries in ${\bf B}_c \to {\bf B}_n M$ were not  discussed in the previous studies 
with $SU(3)_F$ in Refs.~\cite{Lu:2016ogy,first,zero,second,third,fourth,fifth,Wang:2017gxe,Geng:2019bfz,Zhao:2018mov}.

In addition,  it has been noticed  that the physical Cabibble-allowed  dominated decay processes of 
${\bf B}_c \to {\bf B}_n K_L^0$ and ${\bf B}_c \to {\bf B}_n K_S^0$ are  the same when the doubly Cabibbo-suppressed contributions 
are taking to be zero~\cite{Wang:2017gxe}.
However, in some of these processes,  the doubly Cabibbo-suppressed transitions are not negligible,   which can be examined by defining
the $K_S^0-K_L^0$ asymmetries between the $K_S^0$ and $K_L^0$ modes~\cite{Wang:2017gxe}  to track the interferences.

In this work, we will systematically analyze the decay processes of ${\bf B}_c \to {\bf B}_n M$ with the $SU(3)_F$ symmetry
with all operators under $SU(3)_F$. 
We will also include the effect of the $\eta-\eta'$ mixing.
There are two different ways to link the amplitudes among the  processes by  $SU(3)_F$. 
The first one is a purely mathematical consideration. By imposing the $SU(3)$ group, we are able to write down the amplitude by tensor contractions. The second one is the diagrammatic approach,
in which one draws down all the possible diagrams for the  decay process with ascertaining 
 that the  amplitude from each diagram shall be the same by interchanging  up, down and strange quarks. Both ways have their own advantages. The tensor method is easier to cooperate with the other symmetry and it allows us to estimate the order of the contribution from the amplitude with the Wilson coefficients. Explicitly, it could cooperate with the $SU(3)$ color symmetry and take account of the strange quark mass as the source of the $SU(3)_F$ symmetry breaking~\cite{Savage:1991wu,third}. 
On the other hand, the diagrammatic approach can distinguish the factorizable and non-factorizable amplitudes~\cite{Chen:quark}. The close relations between the two methods have been examined in Ref.~\cite{He:2018joe}. 
In Ref.~\cite{fifth}, it has been proved to be useful if one  combines   both methods.

This paper is organized as follows. In Sec. II, we  give the formalism for the two-body charmed baryon decays of ${\bf B}_c \to {\bf B}_n M$, 
in which we first write the decay amplitudes in terms of parity conserved and violated parts under the $SU(3)_F$ flavor symmetry, and  then display the decay rates and asymmetries. 
In Sec.~III, we show our numerical results and  present discussions.
 We conclude  in Sec.~IV. 
 In Appendix A, we list the all decay amplitudes of the anti triplet baryon states in terms of the $SU(3)_F$ parameters.
 We give the definitions of the up-down and longitudinal polarization asymmetries in Appendix~B.
 
\section{Formalism}
To study the two-body decays of the anti-triplet charmed baryon (${\bf B}_c$) to octet baryon  (${\bf B}_n$) and nonet pseudoscalar
meson ($M$) states, 
we write the hadronic state representations under the $SU(3)_F$ flavor symmetry to be
\begin{eqnarray}
&&{\bf B}_{c} = (\Xi_c^0,-\Xi_c^+,\Lambda_c^+)\,,
\nonumber\\
&& {\bf B}_n=\left(\begin{array}{ccc}
\frac{1}{\sqrt{6}}\Lambda+\frac{1}{\sqrt{2}}\Sigma^0 & \Sigma^+ & p\\
 \Sigma^- &\frac{1}{\sqrt{6}}\Lambda -\frac{1}{\sqrt{2}}\Sigma^0  & n\\
 \Xi^- & \Xi^0 &-\sqrt{\frac{2}{3}}\Lambda
\end{array}\right)\,,
\nonumber\\
&&M=\left(\begin{array}{ccc}
\frac{1}{\sqrt{2}}(\pi^0+c_\phi\eta + s_\phi \eta')  & \pi^+ & K^+\\
 \pi^- &\frac{1}{\sqrt{2}}(-\pi^0+c_\phi\eta + s_\phi \eta') &  K^0\\
 K^- & \bar K^0&-s_\phi\eta+  c_\phi \eta'
\end{array}\right)\,,
\end{eqnarray}
respectively, where  $(c_\phi, s_\phi)=(\cos\phi, \sin\phi)$ and $\phi=39.3^\circ$~\cite{FKS} to describe the mixing between $\eta_8$ and $\eta_0$
of the octet and nonet sates for $\eta$.

From  $c\to u \bar{d}s$, $c\to u$ and $c\to u \bar{s}d$ transitions at tree level, the effective Hamiltonian is given by~\cite{Buras:1998raa}
\begin{eqnarray}\label{Heff}
{\cal H}_{eff}&=&\sum_{i=+,-}\frac{G_F}{\sqrt 2}c_i
\left(V_{cs}V_{ud}O^{ds}_i+V_{cd}V_{ud} O^{qq}_i+V_{cd}V_{us}O^{sd}_i\right),
\end{eqnarray}
with
\begin{eqnarray}
\label{O12}
O_\pm^{q_2q_1}&=&{1\over 2}\left[(\bar u q_1)_{V-A}(\bar{q}_2c)_{V-A}\pm (\bar{q}_2 q_1)_{V-A}(\bar u c)_{V-A}\right]\,,
\end{eqnarray}
where $(|V_{cs}V_{ud}|,|V_{cd}V_{ud}|,|V_{cd}V_{us}|)\simeq (1,s_c,s_c^2)$  with $s_c\equiv \sin\theta_c\approx 0.225$~\cite{pdg} 
and $\theta_c$ the Cabibbo angle, $c_i$ (i=+,-) represent the Wilson coefficients, $G_F$ is the Fermi constant, 
  $O_\pm^{q_2q_1}$ and $O_\pm^{qq}\equiv O_\pm^{dd}-O_\pm^{ss}$ are the four-quark operators, and $(\bar q_1 q_2)\equiv\bar q_1\gamma_\mu(1-\gamma_5)q_2$.
In Eq.~(\ref{Heff}),  the decays associated with $O_\pm^{ds}$, $O_\pm^{qq}$ and $O_\pm^{sd}$ are the so-called Cabibbo-allowed (favored), 
singly Cabibbo-suppressed and doubly Cabibbo-suppressed processes, respectively.

 Note that $O_{+(-)}$, corresponding  to the ${\cal O}(\overline{15}(6))$ representation, is (anti)symmetric in flavor and color indices.
 The tensor forms of $H(\overline{15})$ and  $H(6)$ under  $SU(3)_F$ are   given by
\begin{eqnarray}
H(\overline{15})^{ij}_k&=&
\left(\begin{array}{ccc}
\left(\begin{array}{ccc}
0&0&0\\
0&0&0\\
0&0&0
\end{array}\right),
\left(\begin{array}{ccc}
0&s_c&1\\
s_c&0&0\\
1&0&0
\end{array}\right),
\left(\begin{array}{ccc}
0&-s_c^2&-s_c\\
-s_c^2&0&0\\
-s_c&0&0
\end{array}\right)
\end{array}\right)\,,
\nonumber\\
H(6)_{ij}&=&\left(
\begin{array}{ccc}
0& 0 & 0\\
0 & 2 & -2s_c\\
0 & -2s_c& 2s_c^2
\end{array}\right)\,,
\end{eqnarray}
respectively, where we have used the conversion of  $V_{cd}=-V_{us}=s_c$. 
In general, we write the spin-dependent  amplitude of  ${\bf B}_c \to {\bf B}_n M$  as
\begin{equation}
\label{AB}
{\cal M}({\bf B}_c \to {\bf B}_n M)= i \overline{u}_{{\bf B}_n}\left(A-B\gamma_5\right)u_{{\bf B}_c}\,,
\end{equation}
where $A$ and $B$ are the $s$-wave and $p$-wave amplitudes, corresponding to the parity violating and conserving ones,
and $u_{{\bf B}_{n,c}}$ are the baryon Dirac spinors, respectively.
 From Eqs.~(\ref{Heff}) and (\ref{AB}), we can decompose   $A$ in terms of the tensor forms under  $SU(3)_F$ as
\begin{eqnarray}\label{abamp}
&&A_{({\bf B}_c\to {\bf B}_n M)}=\nonumber\\
&& a_0 H(6)_{ij}({\bf B}_c')^{ik}({\bf B}_n)^j_k (M)^l_l 
+a_1 H(6)_{ij}({\bf B}_c')^{ik}({\bf B}_n)_k^l (M)_l^j+a_2 H(6)_{ij}({\bf B}_c')^{ik}(M)_k^l({\bf B}_n)_l^j +\nonumber\\
&&a_3 H(6)_{ij}({\bf B}_n)_k^i (M)_l^j ({\bf B}_c')^{kl}
+a_0'({\bf B}_n)^i_j (M)^l_l H(\overline{15})^{jk}_i({\bf B}_{c})_k
+a_4H(\overline{15})_{k}^{li}({\bf B}_{c})_j (M)_i^j ({\bf B}_n)_l^k+
\nonumber\\
&&
a_5({\bf B}_n)^i_j (M)^l_i H(\overline{15})^{jk}_l ({\bf B}_{c})_k
+a_6({\bf B}_n)^j_i (M)^m_l H(\overline{15})^{li}_m ({\bf B}_{c})_j
+a_7({\bf B}_n)^l_i (M)^i_j H(\overline{15})^{jk}_l ({\bf B}_{c})_k,~~
\nonumber\\
&& 
 B_{({\bf B}_c \to {\bf B}_n M)} = A_{({\bf B}_c\to {\bf B}_n M)}\{a_i^{(\prime)} \rightarrow b_i^{(\prime)}\}
\end{eqnarray}
 where 
 $({\bf B}_c')^{ij}\equiv\epsilon^{ijk}({\bf B}_c)_k$.
Here, we have assumed that the mass dependence of $A$ and $B$ are negligible, while  the Wilson coefficients of $c_i$ have been  absorbed into
 the $SU(3)_F$ parameters $a_i^{(\prime)}$ and $b_i^{(\prime)}$. 
 Note that we treat the $SU(3)_F$ flavor symmetry to be exact. To obtain more precise results, one has to include the $SU(3)_F$ breaking terms in the amplitudes as shown in Refs.~\cite{Savage:1991wu,third}. Note that the analysis with $SU(3)_F$ breaking effect can be done when more experimental data 
 are available  in the future.
 The
  expansions of $A_{({\bf B}_c \to {\bf B}_n M)}$ are listed in Appendix A, while those of $B_{({\bf B}_c \to {\bf B}_n M)}$
 can be derived by replacing $a_i$ in $A_{({\bf B}_c \to {\bf B}_n M)}$ with $b_i$. 
 
 Since the operator ${\cal O}(\overline{15})\sim (\bar u q_1)(\bar{q}_2c)+ (\bar{q}_2 q_1)(\bar u c)$ is symmetric in color index,
whereas  the baryon states  are  antisymmetric, 
the contributions of ${\cal O}(\overline{15})$ from the nonfactorizable part to the amplitude vanish,
so that we only need to consider the factorizable amplitude from ${\cal O}(\overline{15})$~\cite{fifth}.
\begin{figure}[!t]
	\includegraphics[width=5cm]{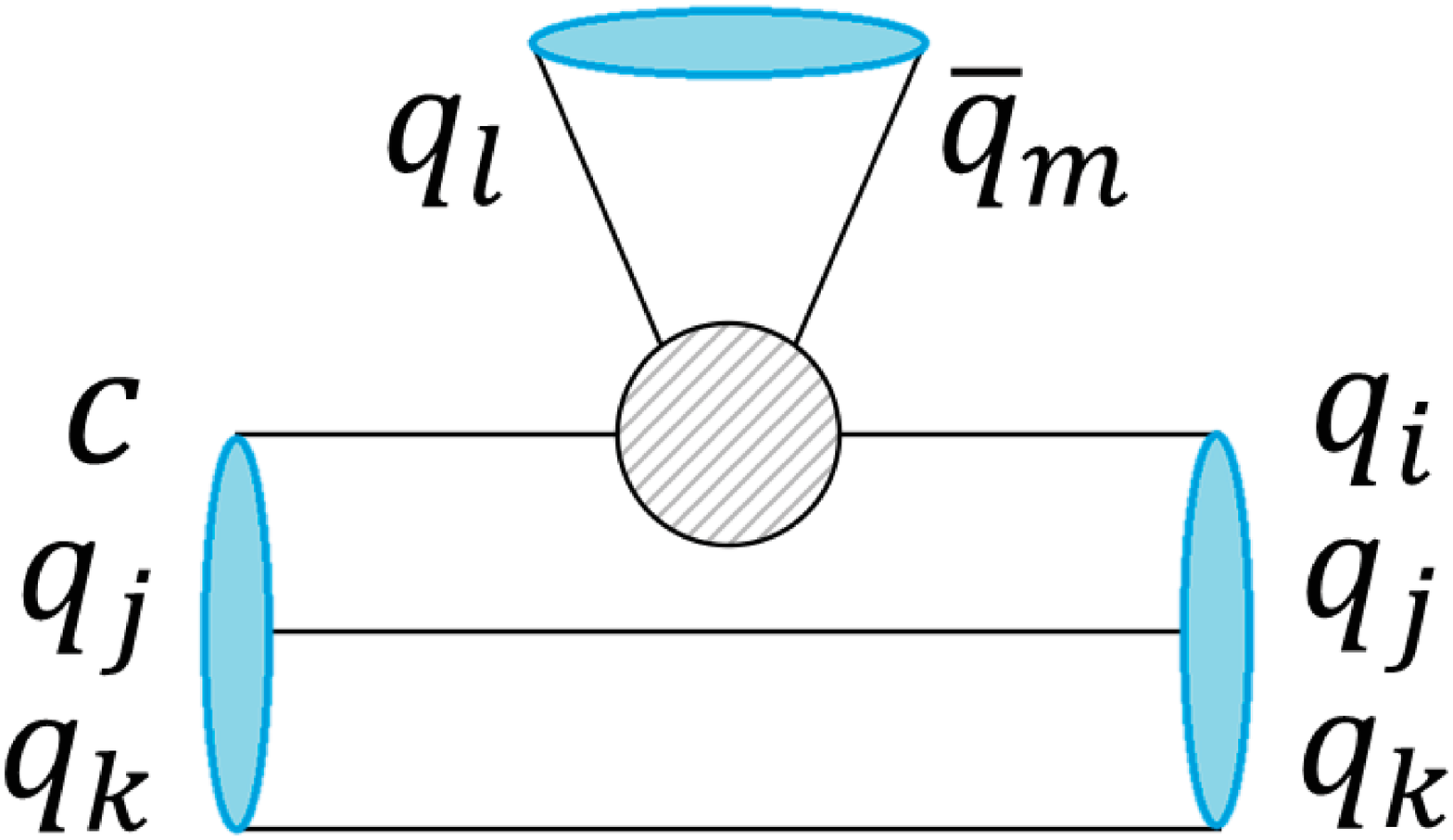}
	\caption{Topological diagram related to factorizable processes with the bubble representing the four-quark interaction.}
	\label{Fig1}
\end{figure}
The factorizable diagram is shown in Fig.~\ref{Fig1} with the bubble representing the four-quark interaction,
which corresponds to 
the factorized amplitude, given by~\cite{Korner92}
\begin{equation}
\frac{G_F}{\sqrt{2}}V_{cq}V_{uq_m}\chi_\pm\langle{\bf B}_n| (\bar{q}_i c)_{V-A}|{\bf B}_c\rangle\langle M| (\bar{q}_l q_m)_{V-A}|0\rangle\,,
\end{equation}
where $q=q_i(q_l)$ and $\chi_\pm$ are related to the effective Wilson coefficients 
 for the charged (neutral) meson in the final states. 

From the topological diagram  in Fig.~\ref{Fig1}, 
one  concludes that only $a_6$ and $b_6$ terms in Eq.~(\ref{abamp}) 
 contain the factorizable contributions  in ${\cal O}(\overline{15})$,  in which the octet meson state $M$ is directly given by the weak interaction alone
 as demonstrated in Ref.~\cite{fifth}.
 As a result, in our calculations we will neglect the terms associated with
 $a_0^{\prime}$, $a_4$, $a_5$ and $a_7$ and  $b_0^{\prime}$, $b_4$, $b_5$ and $b_7$ in Eq.~(\ref{AB}).

 The decay angular distribution of the direction $\hat{p}_{{\bf B}_n}=\vec{p}_{{\bf B}_n}/{p}_{{\bf B}_n}$ 
 ($p_{{\bf B}_n}\equiv |\vec{p}_{{\bf B}_n}|$) of  ${\bf B}_n$ in the rest frame of ${\bf B}_c$ is found to be
 \begin{eqnarray}\label{Angular}
 {d\Gamma \over d\theta}
& \propto & 1+\alpha \vec{P}_{{\bf B}_n}\cdot \hat{p}_{{\bf B}_n}=1+\alpha\cos\theta, 
 \end{eqnarray}
 where $\vec{P}_{{\bf B}_n}$ is the polarization vector of ${\bf B}_n$ with the longitudinal component being $P_{{\bf B}_n}=\alpha$, 
 $\theta$ is the angle between $\vec{P}_{{\bf B}_n}$ and $\hat{p}_{{\bf B}_n}$
 and $\alpha$ is the so-called  up-down asymmetry parameter, given by
\begin{eqnarray}
 \label{Up-daownA}
  \alpha &=& \frac{2\kappa \text{ Re}(A^*B)}{|A|^2+\kappa^2|B|^2}\,,\;\;
   \kappa = \frac{p_{{\bf B}_n}}{E_{{\bf B}_n}+m_{{\bf B}_n}} 
 \end{eqnarray} 
 with $E_{{\bf B}_n}$ and $\vec{p}_{{\bf B}_n}$  the energy and three momentum  of ${\bf B}_n$. 
 The definitions of the up-down and longitudinal asymmetries can be found in Appendix B.
  Consequently, we obtain the decay rate as
\begin{eqnarray}\label{phys}
 \Gamma &=&\frac{p_{{\bf B}_n}}{8\pi}\left(\frac{(m_{{\bf B}_c}+m_{{\bf B}_n})^2-m_M^2}{m_{{\bf B}_c}^2}|A|^2
 +\frac{(m_{{\bf B}_c}-m_{{\bf B}_n})^2-m_M^2}{m_{{\bf B}_c}^2}|B|^2\right)
\end{eqnarray}
To extract the doubly Cabibbo-suppressed contributions in the Cabibbo-allowed dominating decays of  ${\bf B}_c \to {\bf B}_n K_L^0/K_S^0$,
we also define the $K_S^0-K_L^0$ asymmetry parameter as~\cite{Wang:2017gxe} 
\begin{equation}\label{kaonrela}
{\bf R}_{K^0_{S,L}}({\bf B}_c\to {\bf B}_n)=\frac{\Gamma({\bf B}_c\to {\bf B}_n K^0_{S})-\Gamma({\bf B}_c\to {\bf B}_n K^0_{L})}{\Gamma({\bf B}_c\to {\bf B}_n K^0_{S})+\Gamma({\bf B}_c\to {\bf B}_n K^0_{L})}\,.
\end{equation}

\section{Numerical Results and Discussions}

We now determine the $SU3)_F$ parameters through the experimental data~\cite{pdg,Ablikim:2018bir,Exp:eta',Exp:absXic0,YKC}, 
listed in Table~\ref{Table1}, where we have also shown the  reproduced values for the observables. 
In the following analysis, we take the amplitudes of $A$ and $B$ as real by using the fact that CP is mainly conserved in charmed decays and assuming the final state interaction is negligible~\cite{Pakvasa:1990if}\footnote{We note that  $A$ and $B$ are relative real if $CP$ is conserved and the final state interactions are negligible. This statement has been given in many textbooks, such as those in Refs.~\cite{Weinberg:1995mt,Sozzi:2008zza}.}.
Note that in our fit, we have used the original data point of ${\cal B}(\Lambda_c^+\to p\pi^0)=(0.8\pm 1.3)\times 10^{-4}$ from the BESIII Collaboration~\cite{YKC}, but the result of $\alpha(\Lambda_{c}^{+}  \to  \Xi^{0} K^{+})=0.77\pm 0.78$~\cite{Ablikim:2018bir} is not included.
\begin{table}
\caption{Comparisons of the decay branching ratios and asymmetries between the experimental data~\cite{pdg,Ablikim:2018bir,Exp:eta',Exp:absXic0,YKC} 
and theoretical reproductions with $SU(3)_F$.
}
\begin{center}
\begin{tabular}[t]{ccc|cc}
\hline
Channel &${\cal B}_{exp}$ &$\alpha_{exp}$&${\cal B}_{SU(3)_F}$&$\alpha_{SU(3)_F}$\\
\hline
$ \Lambda_{c}^{+}  \to  \Lambda^{0} \pi^{+} $&$(13.0\pm0.7)\times 10^{-3}$&$-0.91\pm 0.15$&$(13.0\pm0.7)\times 10^{-3}$&$-0.87\pm0.10$\\
$ \Lambda_{c}^{+}  \to  p K_S^0 $ &$(15.8\pm0.8)\times 10^{-3}$&&$(15.7\pm0.8)\times 10^{-3}$&$-0.89^{+0.26}_{-0.11}$\\
$ \Lambda_{c}^{+}  \to  \Sigma^{0} \pi^{+} $ &$(12.9\pm0.7)\times 10^{-3}$&&$(12.7\pm0.6)\times 10^{-3}$&$-0.35\pm 0.27$\\
$ \Lambda_{c}^{+}  \to  \Sigma^{+} \pi^{0} $ &$(12.4\pm1.0)\times 10^{-3}$&$-0.45\pm 0.32$&$(12.7\pm 0.6)\times 10^{-3}$&$-0.35\pm 0.27$\\
$ \Lambda_{c}^{+}  \to  \Sigma^{+} \eta $ &$(4.1\pm2.0)\times 10^{-3}$&&$(3.2\pm1.3)\times 10^{-3}$&$-0.40\pm0.47$\\
$ \Lambda_{c}^{+}  \to  \Sigma^{+} \eta' $ &$(13.4\pm5.7)\times 10^{-3}$&&$(14.4\pm5.6)\times 10^{-3}$&$1.00^{+0.00}_{-0.17}$\\
$ \Lambda_{c}^{+}  \to  \Xi^{0} K^{+} $ &  $(5.9\pm1.0)\times 10^{-3}$&$^*0.77\pm0.78$&$(5.6\pm0.9)\times 10^{-3}$&$0.94^{+0.06}_{-0.11}$\\
\hline
$ \Lambda_{c}^{+}  \to  p \pi^{0} $ &$(0.8\pm1.3)\times 10^{-4}$\cite{YKC}&&$(1.2\pm1.2)\times 10^{-4}$&$-0.05\pm0.72$\\
$ \Lambda_{c}^{+}  \to  p \eta $ &$(12.4\pm3.0)\times 10^{-4}$&&$(11.5\pm2.7)\times 10^{-4}$&$-0.96^{+0.30}_{-0.04}$\\
$ \Lambda_{c}^{+}  \to  \Lambda^{0} K^{+} $ &$(6.1\pm1.2)\times 10^{-4}$&&$(6.5\pm1.0)\times 10^{-4}$&$0.32\pm0.30$\\
$ \Lambda_{c}^{+}  \to  \Sigma^{0} K^{+} $ &$(5.2\pm0.8)\times 10^{-4}$&&$(5.4\pm0.7)\times 10^{-4}$&$-1.00^{+0.06}_{-0.00}$\\
\hline
$ \Xi_{c}^{0}  \to  \Xi^{-} \pi^{+} $ &$ (1.80 \pm 0.52)\times 10^{-2}$ &$-0.6\pm0.4$&$ (2.21 \pm 0.14)\times 10^{-2}$&$-0.98^{+0.07}_{-0.02}$ \\
$ \Xi_{c}^{0}  \to  \Lambda^{0} K_S^0 $ &&&$ (5.0 \pm 0.3)\times 10^{-3}$&$-0.70\pm 0.28$\\
$^{**}{\cal R}_{\Xi_c^0}$ & $0.210\pm0.028$&\\
\hline
\label{Table1}
\end{tabular}
\end{center}
 $^*$This value is not included in the data input.~~ $^{**}{\cal R}_{\Xi_c^0}\equiv {\cal B}(\Xi_c^0 \to \Lambda K_S^0)/{\cal B}(\Xi_c^0\to\Xi^-\pi^+ )$.
\end{table}
Consequently,
there are  16 experimental data  inputs to fit with 10 $SU(3)_F$ parameters in Eq.~(\ref{abamp}), given by
\begin{equation}\label{subse}
(a_1,a_2,a_3,a_6,\tilde{a},b_1,b_2,b_3,b_6,\tilde{b})\,,
\end{equation}
resulting in the degree of freedom (d.o.f) to be 6. 
In order to  separate  the amplitudes from $\eta_0$ and octet meson states,
we  define $\tilde{a}$ and $\tilde{b}$ by
\begin{eqnarray}
\tilde{a}\equiv a_0 +\frac{1}{3}\left(a_1+a_2-a_3\right)\,,~~
 \tilde{b}\equiv b_0 +\frac{1}{3}\left(b_1+b_2-b_3\right)
 \end{eqnarray}
 respectively.
As a result, the $\eta_0$ amplitude depends only on $\tilde{a}$ and $\tilde{b}$. 
By performing the minimal $\chi^2$ fitting  as shown in Ref.~\cite{second}, we obtain
{\footnotesize
\begin{eqnarray}\label{fittingresults}
  (a_1,a_2,a_3,a_6,\tilde{a}) 
   =(4.34\pm0.50,-1.33\pm0.32,1.25\pm0.36,-0.26\pm0.64,1.77\pm0.83)\, 10^{-2}G_F\text{GeV}^2 ,\,
  \nonumber \\
  (b_1,b_2,b_3,b_6,\tilde{b})
    =(-9.20\pm2.09,-8.03\pm1.19,1.42\pm1.61,-4.05\pm2.48,13.15\pm5.56)\, 10^{-2}G_F\text{GeV}^2.\,
\end{eqnarray}
}
The  correlation coefficients between i-th and j-th $SU(3)_F$ parameters in Eq.~(\ref{subse}) are given by
{\small
\begin{eqnarray}\label{Correl}
R&=&\left(
\begin{array}{cccccccccc}
1& 0.64& 0.59& -0.58& -0.30& 0.96& -0.47& 0.67& -0.66& 0.25\\
0.64& 1& -0.17& -0.07& -0.38& 0.61& -0.59& 0.38& -0.12& 0.36\\
0.59& -0.17& 1& -0.67& 0.01& 0.55& 0.03& 0.57& -0.75& -0.05\\
-0.58& -0.07& -0.67& 1& 0.11& -0.65& 0.21& -0.59& 0.93& -0.05\\
-0.30& -0.38& 0.01& 0.11& 1& -0.31& 0.34& -0.29& 0.15& -0.35\\
0.96& 0.61& 0.55& -0.65& -0.31& 1& -0.51& 0.63& -0.70& 0.27\\
-0.47& -0.59& 0.03& 0.21& 0.34& -0.51& 1& -0.59& 0.15& -0.29\\
0.67& 0.38& 0.57& -0.59& -0.29& 0.63& -0.59& 1& -0.69& 0.22\\
-0.66& -0.12& -0.75& 0.93& 0.15& -0.70& 0.15& -0.69& 1& -0.10\\
0.25& 0.36& -0.05& -0.05& -0.35& 0.27& -0.29& 0.22& -0.10& 1
\end{array}\right)\,.
\end{eqnarray}
}
In our fit, we find that $\chi^2/d.o.f=0.5$, which indicates that our results with the $SU(3)_F$ symmetry can well
explain all current existing experimental data for the decay branching ratios and  up-down asymmetries. 
Indeed, as seen in Table~\ref{Table1}, our reproductions based on $SU(3)_F$ are all consistent with  the corresponding experimental measurements.
However, it is important to pointed out that our values of  ${\cal B}( \Xi_{c}^{0}  \to  \Xi^{-} \pi^{+})=(2.21\pm0.14)\times 10^{-2} $
and $|\alpha( \Xi_{c}^{0}  \to  \Xi^{-} \pi^{+})|=0.98^{+0.02}_{-0.07}$ are consistent with, but higher than, the corresponding data of
$(1.80\pm0.52)\times 10^{-2} $~\cite{Exp:absXic0}
 and $0.6\pm0.4$~\cite{pdg}, respectively.

It is worth to take a closer look on the parameters  in Eq.~(\ref{fittingresults}).
As mentioned early, $H(\overline{15})$ only contributes to the factorization amplitudes, which can be  parametrized only
in terms of
 $a_6$ and $b_6$ terms, corresponding to the vector and axial-vector currents in the baryonic matrix elements, respectively. 
 Our result of $b_6\gg a_6$ in Eq.~(\ref{fittingresults}) suggests that
  the axial-vector  part of the factorization contribution 
  is much larger that the vector one. This can be understood as follows.
 In the  decay of ${\bf B}_c\to {\bf B}_n M$,   the pseudoscalar meson part of the factorization approach is given by
\begin{equation}\label{fact}
  \langle0|j^\mu_5|M\rangle=if_Mq^\mu\,,
\end{equation}
where $f_M$ is   the  meson decay constant, while
 $q^\mu$ is the four-momentum of $M$, which is also equal to 
 the four-momentum difference between  the initial and final baryons of ${\bf B}_c$ and ${\bf B}_n$.
 Consequently, we get that 
\begin{equation}\label{facre}
q^\mu\langle B_n|\bar{q}\gamma_\mu \gamma_5 c|B_c\rangle\gg q^\mu\langle B_n|\bar{q}\gamma_\mu c|B_c\rangle=i\langle B_n|\partial^\mu(\bar{q}\gamma_\mu c)|B_c\rangle \,,
\end{equation}
 where $q$ stands for the light quarks.
 In the case of the $SU(4)$ flavor symmetry, in which the charm quark is also treated as $q$, 
Eq.~\eqref{facre} is automatically satisfied as the right-handed part is zero.
It is clear that the inequality in Eq.~(\ref{facre})  depends on the parameters $a_6$ and $b_6$, which are not quite determined yet, particularly 
$a_6$.
In fact, from Table~\ref{Table7} in Appendix~A, we have that
\begin{equation}\label{ppi0}
A(\Lambda_c^+ \to p \pi^0)=\sqrt{2}\left(a_2+a_3-\frac{a_6}{2}\right)\,,
\end{equation}
in which $a_2$ and $a_3$ get almost canceled out each other, resulting in that it could be dominated by the $a_6$ terms.
In this case,  the experimental search for the up-down asymmetry as well as the future measurement on the branching ration of
 $\Lambda_c^+ \to p \pi^0$ will be helpful to obtain  the precise value of $a_6$.

In Tables~\ref{Table2}, \ref{Table3} and \ref{Table4},
 we list  our predictions of the  branching ratios and up-down asymmetries for the
 Cabibbo-allowed, singly Cabibbo-suppressed and doubly Cabibbo-suppressed decays, respectively. 
 In the tables, we have also presented  the values of A and B, which are useful to understand the up-down asymmetries as well as the comparisons 
 with those given by specific  theoretical models. We note that some of our results for the up-down asymmetries have been discussed 
 for the first time in the literature, while the  decay branching  ratios are  almost the same as those
 in Refs.~\cite{Lu:2016ogy,first,zero,second,third,fifth,Wang:2017gxe}. In particular, we find that
 ${\cal B}(\Lambda_c^+\to p \pi^0)=(1.2\pm1.2)\times 10^{-4}$, which is consistent with our previous value of $(1.3\pm0.7)\times 10^{-4}$ in Ref.~\cite{fifth}
 and $0.8\times 10^{-4}$ calculated by the pole model with current algebra in Ref.~\cite{Cheng:sup} as well as 
   the current experimental upper limit of $2.7\times 10^{-4}$~\cite{pdg}. 
   In addition, the decay branching ratio for the related Cabibbo-suppressed mode of
   $\Lambda_c^+\to n \pi^+$ is predicted to be $(8.5\pm1.9)\times 10^{-4}$, in comparison with $(6.1\pm2.0)\times 10^{-4}$ in Ref.~\cite{fifth}
   and $2.7\times 10^{-4}$ in Ref.~\cite{Cheng:sup}. We remark that most of the branching ratios in the present work with the spin-dependent amplitudes
   have small uncertainties comparing to those of our previous study with $SU(3)_F$ in Ref.~\cite{fifth} except the decay of  $\Lambda_c^+\to p \pi^0$
   due to the cancellation effect  as well as the correlations in Eq.~(\ref{Correl}).
   Explicitly, as shown Table~III,
   the sign in $A_{(\Lambda_c^+ \to p \pi^0)} = (-0.01\pm0.10)\text{sin}\theta_c\times10^{-1}G_F\text{GeV}^2$
    is  not well determined, resulting in a large  error in $\alpha(\Lambda_c^+ \to p \pi^0)_{SU(3)}=-0.05\pm0.72$.
   To determine the asymmetry precisely, the experiment with a smaller uncertainty is clearly needed.

\begin{table}
\begin{center}
\caption{Predictions of the  branching ratios and up-down asymmetries for the
 Cabibbo-allowed decays, where we have also listed the values of A and B in the unit of $10^{-1}G_F\text{GeV}^2$.}\label{Table2}
\setlength{\tabcolsep}{5mm}{
\begin{tabular}[t]{lcccc}
\hline
channel & $A$&$B $&$10^3{\cal B}$& $\alpha$\\
\hline
$ \Lambda_{c}^{+}  \to  \Lambda^{0} \pi^{+} $&$ -0.33 \pm 0.06 $&$ 1.62 \pm 0.12 $&$ 13.0 \pm 0.7 $&$ -0.87 \pm 0.10$\\
$ \Lambda_{c}^{+}  \to  p \bar{K}^{0} $&$ -0.89 \pm 0.15 $&$ 1.44 \pm 0.62 $&$ 31.2 \pm 1.6 $&$ -0.90 ^{ +0.22}_{-0.10} $\\
$ \Lambda_{c}^{+}  \to  \Sigma^{0} \pi^{+} $&$ -0.63 \pm 0.02 $&$ 0.37 \pm 0.29 $&$ 12.7 \pm 0.6 $&$ -0.35 \pm 0.27$\\
$ \Lambda_{c}^{+}  \to  \Sigma^{+} \pi^{0} $&$ 0.63 \pm 0.02 $&$ -0.37 \pm 0.29 $&$ 12.7 \pm 0.6 $&$ -0.35 \pm 0.27$\\
$ \Lambda_{c}^{+}  \to  \Sigma^{+} \eta $&$ -0.34 \pm 0.07 $&$ 0.26 \pm 0.44 $&$ 3.2 \pm 1.3 $&$ -0.40 \pm  0.47 $\\
$ \Lambda_{c}^{+}  \to  \Sigma^{+} \eta' $&$ -0.69 \pm 0.26 $&$ -4.80 \pm 1.54 $&$ 14.4 \pm 5.6 $&$ 1.00 ^{+ 0.00 }_{- 0.17 }$\\
$ \Lambda_{c}^{+}  \to  \Xi^{0} K^{+} $&$ 0.27 \pm 0.06 $&$ 1.61 \pm 0.24 $&$ 5.6 \pm 0.9 $&$ 0.94 ^{+0.06}_{-0.11}$\\
\hline
$ \Xi_{c}^{+}  \to  \Sigma^{+} \bar{K}^{0} $&$ 0.28 \pm 0.12 $&$ 0.69 \pm 0.52 $&$ 8.6^{+9.4}_{-7.8} $&$ 0.98 _{- 0.16}^{+0.02} $\\
$ \Xi_{c}^{+}  \to  \Xi^{0} \pi^{+} $&$ -0.22 \pm 0.06 $&$ 0.12 \pm 0.23 $&$ 3.8 \pm 2.0 $&$ -0.32 \pm 0.52$\\
\hline
$ \Xi_{c}^{0}  \to  \Xi^{-} \pi^{+} $&$ 0.84 \pm 0.08 $&$ -2.25 \pm 0.3 $&$ 22.1 \pm 1.4 $&$ -0.98 ^{+ 0.07 }_{- 0.02 }$\\
$ \Xi_{c}^{0}  \to  \Lambda^{0} \bar{K}^{0} $&$ -0.73 \pm 0.07 $&$ 0.80 \pm 0.39 $&$ 10.5 \pm 0.6 $&$ -0.68 \pm 0.28 $\\
$ \Xi_{c}^{0}  \to  \Sigma^{0} \bar{K}^{0} $&$ -0.01 \pm 0.10 $&$ 0.65 \pm 0.33 $&$ 0.8 \pm 0.8 $&$ -0.07 \pm 0.90 $\\
$ \Xi_{c}^{0}  \to  \Sigma^{+} K^{-} $&$ -0.27 \pm 0.06 $&$ -1.61 \pm 0.24 $&$ 5.9 \pm 1.1 $&$ 0.81 \pm 0.16$\\
$ \Xi_{c}^{0}  \to  \Xi^{0} \pi^{0} $&$ -0.44 \pm 0.06 $&$ 1.50 \pm 0.23 $&$ 7.6 \pm 1.0 $&$ -1.00^{+ 0.07 }_{- 0.00 }$\\
$ \Xi_{c}^{0}  \to  \Xi^{0} \eta $&$ 0.65 \pm 0.08 $&$ 1.64 \pm 0.55 $&$ 10.3 \pm 2.0 $&$ 0.93 ^{+ 0.07 }_{- 0.19 }$\\
$ \Xi_{c}^{0}  \to  \Xi^{0} \eta' $&$ 0.61 \pm 0.25 $&$ 4.27 \pm 1.51 $&$ 9.1 \pm 4.1 $&$ 0.98 ^{+ 0.02 }_{- 0.27 }$\\
\hline
\end{tabular}
}
\end{center}
\end{table}
\renewcommand\arraystretch{0.9}
\begin{table}
\begin{center}
\centering
\caption{Legend is the same as Table~\ref{Table2} but for the singly Cabibbo-suppressed decays with an overall factor of $\sin\theta_c$ for
$A$ and $B$ omitted.}\label{Table3}
\setlength{\tabcolsep}{5mm}{
\begin{tabular}[t]{lcccc}
\hline
channel & $A$&$B $&$10^4{\cal B}$& $\alpha$\\
\hline
$ \Lambda_{c}^{+}  \to  p \pi^{0} $&$ 0.01 \pm 0.10 $&$ -0.65 \pm 0.33 $&$ 1.2 \pm 1.2 $&$ -0.05 \pm 0.72 $\\
$ \Lambda_{c}^{+}  \to  p \eta $&$ -0.75 \pm 0.18 $&$ 1.44 \pm 0.77 $&$ 12.4 \pm 3.5 $&$ -0.94 ^{+0.26}_{-0.06} $\\
$ \Lambda_{c}^{+}  \to  p \eta' $&$ 0.84 \pm 0.27 $&$ 4.33 \pm 1.91 $&$ 24.5 \pm 14.6 $&$ 0.91 _{-0.21}^{+0.09} $\\
$ \Lambda_{c}^{+}  \to  n \pi^{+} $&$ -0.04 \pm 0.07 $&$ -1.73 \pm 0.20 $&$ 8.5 \pm 2.0 $&$ 0.12 \pm 0.19 $\\

$ \Lambda_{c}^{+}  \to  \Lambda^{0} K^{+} $&$ 0.65 \pm 0.06 $&$ 0.35 \pm 0.35 $&$ 6.5 \pm 1.0 $&$ 0.32 \pm 0.32 $\\
$ \Lambda_{c}^{+}  \to  \Sigma^{0} K^{+} $&$ 0.44 \pm 0.06 $&$ -1.50 \pm 0.23 $&$ 5.4 \pm 0.7 $&$ -1.00 ^{+0.06}_{-0.00} $\\
$ \Lambda_{c}^{+}  \to  \Sigma^{+} K^{0} $&$ 0.62 \pm 0.08 $&$ -2.12 \pm 0.33 $&$ 10.9 \pm 1.5 $&$ -1.0  ^{+0.06}_{-0.00} $\\
\hline
$ \Xi_{c}^{+}  \to  \Lambda^{0} \pi^{+} $&$ 0.05 \pm 0.07 $&$ -1.47 \pm 0.24 $&$ 12.3 \pm 4.1 $&$ -0.19 \pm 0.24 $\\
$ \Xi_{c}^{+}  \to  p \bar{K}^{0} $&$ 0.62 \pm 0.08 $&$ -2.12 \pm 0.33 $&$ 43.3 \pm 7.8 $&$ -0.93 ^{+0.09}_{-0.07}$\\
$ \Xi_{c}^{+}  \to  \Sigma^{0} \pi^{+} $&$ 0.78 \pm 0.04 $&$ -0.45 \pm 0.34 $&$ 25.5 \pm 2.6 $&$ -0.38 \pm 0.27 $\\
$ \Xi_{c}^{+}  \to  \Sigma^{+} \pi^{0} $&$ -0.82 \pm 0.09 $&$ -0.12 \pm 0.53 $&$ 26.9 \pm 6.5 $&$ 0.10 \pm 0.43 $\\
$ \Xi_{c}^{+}  \to  \Sigma^{+} \eta $&$ 0.63 \pm 0.16 $&$ 0.56 \pm 0.85 $&$ 15.5 \pm 10.3 $&$ 0.58^{+0.42}_{-0.59}$\\
$ \Xi_{c}^{+}  \to  \Sigma^{+} \eta' $&$ 0.46 \pm 0.29 $&$ 4.57 \pm 1.84 $&$ 34.6 \pm 21.9 $&$ 0.72 _{-0.41}^{+0.28} $\\
$ \Xi_{c}^{+}  \to  \Xi^{0} K^{+} $&$ -0.04 \pm 0.07 $&$ -1.73 \pm 0.20 $&$ 8.2 \pm 1.9 $&$ 0.17 \pm 0.28 $\\
\hline
$ \Xi_{c}^{0}  \to  \Lambda^{0} \pi^{0} $&$ -0.02 \pm 0.06 $&$ 1.28 \pm 0.21 $&$ 2.3 \pm 0.8 $&$ -0.09 \pm 0.23 $\\
$ \Xi_{c}^{0}  \to  \Lambda^{0} \eta $&$ -0.10 \pm 0.15 $&$ 2.95 \pm 0.65 $&$ 6.4 \pm 2.3 $&$ -0.42 \pm 0.27 $\\
$ \Xi_{c}^{0}  \to  \Lambda^{0} \eta' $&$ 0.81 \pm 0.33 $&$ 4.97 \pm 2.28 $&$ 16.4 \pm 10.6 $&$ 0.87_{- 0.28}^{+0.13} $\\
$ \Xi_{c}^{0}  \to  p K^{-} $&$ 0.27 \pm 0.06 $&$ 1.61 \pm 0.24 $&$ 5.0 \pm 1.1 $&$ 0.67 \pm 0.17 $\\
$ \Xi_{c}^{0}  \to  n \bar{K}^{0} $&$ 0.88 \pm 0.03 $&$ -0.52 \pm 0.42 $&$ 7.5 \pm 0.5 $&$ -0.47 \pm 0.34 $\\
$ \Xi_{c}^{0}  \to  \Sigma^{0} \pi^{0} $&$ 0.31 \pm 0.09 $&$ -1.52 \pm 0.28 $&$ 3.8 \pm 0.7 $&$ -0.88^{+0.19}_{-0.12} $\\
$ \Xi_{c}^{0}  \to  \Sigma^{0} \eta $&$ -0.44 \pm 0.11 $&$ -0.39 \pm 0.60 $&$ 1.4 \pm 0.8 $&$ 0.09 \pm 0.77 $\\
$ \Xi_{c}^{0}  \to  \Sigma^{0} \eta' $&$ -0.33 \pm 0.20 $&$ -3.23 \pm 1.30 $&$ 3.3 \pm 2.2 $&$ 0.70 _{- 0.43}^{+0.30} $\\
$ \Xi_{c}^{0}  \to  \Sigma^{+} \pi^{-} $&$ -0.27 \pm 0.06 $&$ -1.61 \pm 0.24 $&$ 3.9 \pm 0.8 $&$ 0.78 \pm 0.17 $\\
$ \Xi_{c}^{0}  \to  \Sigma^{-} \pi^{+} $&$ 0.84 \pm 0.08 $&$ -2.24 \pm 0.30 $&$ 13.3 \pm 0.9 $&$ -1.00 ^{+0.02}_{-0.00} $\\
$ \Xi_{c}^{0}  \to  \Xi^{0} K^{0} $&$ -0.88 \pm 0.03 $&$ 0.52 \pm 0.42 $&$ 7.2 \pm 0.4 $&$ -0.32 \pm 0.25 $\\
$ \Xi_{c}^{0}  \to  \Xi^{-} K^{+} $&$ -0.84 \pm 0.08 $&$ 2.24 \pm 0.30 $&$ 9.8 \pm 0.6 $&$ -0.95_{-0.05}^{+0.06} $\\
\hline
\end{tabular}
}
\end{center}
\end{table}
\renewcommand\arraystretch{1}
\begin{table}
\begin{center}
\centering
\caption{Legend is the same as Table~\ref{Table2} but for the doubly Cabibbo-suppressed decays with an overall factor of $\sin^2\theta_c$ for
$A$ and $B$ omitted.}\label{Table4}
\setlength{\tabcolsep}{5mm}{
\begin{tabular}[t]{lcccc}
\hline
channel & $A$&$B $&$10^5{\cal B}$& $\alpha$\\
\hline
$ \Lambda_{c}^{+}  \to  p K^{0} $&$ 0.28 \pm 0.12 $&$ 0.69 \pm 0.52 $&$ 1.2 ^{+1.4}_{-1.2} $&$ 1.00^{+0}_{- 0.09} $\\
$ \Lambda_{c}^{+}  \to  n K^{+} $&$ -0.22 \pm 0.06 $&$ 0.12 \pm 0.23 $&$ 0.4 \pm 0.2 $&$ -0.41 ^{+0.62}_{-0.59} $\\
\hline
$ \Xi_{c}^{+}  \to  \Lambda^{0} K^{+} $&$ -0.38 \pm 0.03 $&$ -0.49 \pm 0.24 $&$ 3.3 \pm 0.8 $&$ 0.76 \pm 0.24 $\\
$ \Xi_{c}^{+}  \to  p \pi^{0} $&$ 0.19 \pm 0.05 $&$ 1.14 \pm 0.17 $&$ 6.0 \pm 1.4 $&$ 0.65 \pm 0.17 $\\
$ \Xi_{c}^{+}  \to  p \eta $&$ 0.43 \pm 0.10 $&$ -2.25 \pm 0.53 $&$ 20.4 \pm 8.4 $&$ -0.75 \pm 0.15 $\\
$ \Xi_{c}^{+}  \to  p \eta' $&$ -0.75 \pm 0.27 $&$ -4.10 \pm 1.87 $&$ 40.1 \pm 27.7 $&$ 0.80 ^{+0.20}_{-0.30} $\\
$ \Xi_{c}^{+}  \to  n \pi^{+} $&$ 0.27 \pm 0.06 $&$ 1.61 \pm 0.24 $&$ 12.1 \pm 2.8 $&$ 0.65 \pm 0.17 $\\
$ \Xi_{c}^{+}  \to  \Sigma^{0} K^{+} $&$ -0.60 \pm 0.06 $&$ 1.59 \pm 0.21 $&$ 11.9 \pm 0.7 $&$ -0.99 ^{+0.03}_{-0.01} $\\
$ \Xi_{c}^{+}  \to  \Sigma^{+} K^{0} $&$ -0.89 \pm 0.15 $&$ 1.44 \pm 0.62 $&$ 19.5 \pm 1.7 $&$ -0.82 \pm ^{+0.28}_{-0.18} $\\
\hline
$ \Xi_{c}^{0}  \to  \Lambda^{0} K^{0} $&$ -0.36 \pm 0.05 $&$ -0.16 \pm 0.26 $&$ 0.6 \pm 0.2 $&$ 0.32 \pm 0.45 $\\
$ \Xi_{c}^{0}  \to  p \pi^{-} $&$ 0.27 \pm 0.06 $&$ 1.61 \pm 0.24 $&$ 3.1 \pm 0.7 $&$ 0.65 \pm 0.17 $\\
$ \Xi_{c}^{0}  \to  n \pi^{0} $&$ -0.19 \pm 0.05 $&$ -1.14 \pm 0.17 $&$ 1.5 \pm 0.4 $&$ 0.65 \pm 0.17 $\\
$ \Xi_{c}^{0}  \to  n \eta $&$ 0.43 \pm 0.10 $&$ -2.25 \pm 0.53 $&$ 5.2 \pm 2.1 $&$ -0.75 \pm 0.15 $\\
$ \Xi_{c}^{0}  \to  n \eta' $&$ -0.75 \pm 0.27 $&$ -4.10 \pm 1.87 $&$ 10.2 \pm 7.1 $&$ 0.80 ^{+0.20}_{-0.30} $\\
$ \Xi_{c}^{0}  \to  \Sigma^{0} K^{0} $&$ 0.63 \pm 0.10 $&$ -1.01 \pm 0.44 $&$ 2.5 \pm 0.2 $&$ -0.82 ^{+0.28}_{-0.18}$\\
$ \Xi_{c}^{0}  \to  \Sigma^{-} K^{+} $&$ -0.84 \pm 0.08 $&$ 2.24 \pm 0.30 $&$ 6.1 \pm 0.4 $&$ -0.99^{+0.03}_{-0.01} $\\
\hline
\end{tabular}
}
\end{center}
\end{table}

To compare our predictions of the up-down asymmetries with those in the literature, we summarize 
the values of $\alpha$ for the Cabibbo-allowed and singly Cabibbo-suppressed decays of ${\bf B}_c \to {\bf B}_n M$ in
Tables~\ref{Table_alpha} and \ref{Table_alpha_sup}, respectively. 
\begin{table}
	\begin{center}
		\centering
		\caption{Summary of our results with $SU(3)_F$ and those in the literature for the up-down asymmetries of
			the Cabibno-allowed charmed baryon decays,
			where the data, KK, XK, CT, UVK, Zen, Iva, SV1, and SV2 are from
			the PDG~\cite{pdg},  
			Korner and Kramer~\cite{Korner92}, Xu and Kamal~\cite{Xu92}, Cheng and Tseng~\cite{Cheng:allow}, 
			Uppal, Verma and Khanna~\cite{Uppal:1994pt},
			Zenczykowski~\cite{xialpha2}, Ivanov $el~al.$~\cite{xialpha1},
			Sharma and Verma~\cite{Verma98}, and  Sharma and Verma ~\cite{Sharma:1996sc}, respectively.
		}\label{Table_alpha}
		{
		\scriptsize
			\begin{tabular}[t]{lcccccccccc}
				\hline
				channel & our result & data & KK& XK& CT & UVK&Zen& Iva& SV1 &SV2  \\
				 & &  & &  & (CT$^\prime$) & (UVK$^\prime$) & &  &  & (SV2$^\prime$)  \\
				\hline
				$ \Lambda_{c}^{+}  \to  \Lambda^{0} \pi^{+} $ 
				& $ -0.87 \pm 0.10$ & $-0.91\pm0.15$ 
				&$-0.70$& $-0.67$& $-0.99$& $-0.87 $&$-0.99$&$-0.95$&$-0.99$ &input 
				\\ 
				& & & & & $(-0.95)$& $(-0.85) $&&& & 
				\\
				$ \Lambda_{c}^{+}  \to  p \bar{K}^0 $ 
				&$ -0.90^{+0.22}_{-0.10}$  &&$-1.0$&$0.51$&$-0.90$&$-0.99$&$-0.66$&$-0.97$&$-0.99$&$-0.99\pm0.39$\\ 
				&  &&&&$(-0.49)$&$(-0.99)$&&&&\\
				$ \Lambda_{c}^{+}  \to  \Sigma^{0} \pi^{+} $
				&$ -0.35 \pm 0.27$ &&$0.70$& $0.92$&$-0.49$&$-0.32$&
				$0.39$&$0.43$& $-0.31$&$-0.45\pm0.32$\\
				&&&& &$(0.78)$&$(-0.32)$&&& &\\
				$ \Lambda_{c}^{+}  \to  \Sigma^{+} \pi^{0} $
				&$ -0.35 \pm 0.27$ & $-0.45\pm0.32$ 
				&$0.70$& $0.92$&$-0.49$&$-0.32$& $0.39$&$0.43$&$-0.31$&input\\
				& &  && &$(0.78)$&$(-0.32)$& &&&\\
				$ \Lambda_{c}^{+}  \to  \Sigma^{+} \eta $
				&$ -0.40 \pm  0.47 $ &&$0.33$&&&$-0.94$&0&0.55& $-0.99$&$0.92\pm0.47$\\
				& &&&&&$(-0.99)$&&& &$(0.96\pm0.34)$\\
				$ \Lambda_{c}^{+}  \to  \Sigma^{+} \eta' $
				&$ 1.00 ^{+ 0.00 }_{- 0.17 }$ &&$-0.45$&&&$0.68$& $-0.91$&$-0.05$&$0.44$&$-0.75\pm0.38$ \\
				& &&&&&$(0.68)$& &&$0.44$&$(-0.91\pm0.40)$ \\
				$ \Lambda_{c}^{+}  \to  \Xi^{0} K^{+} $
				&$ 0.94 ^{+0.06}_{-0.11}$ & $0.77\pm 0.78$ 
				&0&  0&& 0&0&0&0&0 \\
				\hline
				$ \Xi_{c}^{+}  \to  \Sigma^{+} \bar{K}^0 $&$ 0.98 ^{+ 0.02 }_{- 0.16 }
				  $&&$-1.0$&$0.24$&$0.43$&&$1.0$&$-0.99$&$-0.38$&$0.03\pm0.31$\\
				&&&&&$(-0.09)$&&&&&$(-0.23\pm0.22)$\\
				$ \Xi_{c}^{+}  \to  \Xi^{0} \pi^{+} $&$ -0.32 \pm 0.52$&&$-0.78$&$-0.81$&$-0.77$& 
				&$1.0$&$-1.0$&$-0.74$&$0.03\pm0.29$\\
                                 &&&&&$(-0.77)$&&&&&$(-0.24\pm0.23)$\\
				\hline
				$ \Xi_{c}^{0}  \to  \Xi^{-} \pi^{+} $ & $ -0.98 ^{+ 0.07 }_{- 0.02 }$ &$-0.6\pm0.4$ 
				&$-0.38$&$-0.38$&$-0.47$&&$-0.79$&$-0.84$&$-0.99$&$-0.96\pm0.38$\\
				&&&&&$(-0.99)$\\
				$ \Xi_{c}^{0}  \to  \Lambda^{0} \bar{K}^0 $&$ -0.68 \pm 0.28$&&$-0.76$&$1.0$&$-0.88$&&$-0.29$&$-0.75$&$-0.85$&$-0.85\pm0.36$\\
				&&&&&$(-0.73)$\\
				$ \Xi_{c}^{0}  \to  \Sigma^{0} \bar{K}^0 $&$ -0.07 \pm 0.90$&&$-0.96$&$-0.99$&$0.85$&&$-0.50$&$-0.55$&$-0.15$&$0.07\pm0.67$\\
				&&&&&$(-0.59)$\\
				$ \Xi_{c}^{0}  \to  \Sigma^{+} K^{-} $&$ 0.81 \pm 0.16$&&0&0&&&0&0&0&0\\
				$ \Xi_{c}^{0}  \to  \Xi^{0} \pi^{0} $&$ -1.00^{+ 0.07 }_{- 0.00 }$&&$0.92$&$0.92$&$-0.78$&&$0.21$&$0.94$&$-0.80$&$-0.99\pm0.37$\\
				&&&&&$(-0.54)$\\
				$ \Xi_{c}^{0}  \to  \Xi^{0} \eta $&$ 0.93 ^{+ 0.07 }_{- 0.19 }$&&$-0.92$&&&&$-0.04$&$-1.0$&$-0.45$&$-0.96\pm0.38$\\
				&&&&&&&&&&$(0.14\pm0.34)$\\
				$ \Xi_{c}^{0}  \to  \Xi^{0} \eta' $&$ 0.98 ^{+ 0.02 }_{- 0.27 }$&&$-0.38$&&&&$-1.0$&$-0.32$&$0.65$&$-0.63\pm0.40$\\
				&&&&&&&&&&$(-0.99\pm0.42)$\\
				\hline
			\end{tabular}
		}
	\end{center}
\end{table}
\begin{table}
	\begin{center}
		\centering
		\caption{Summary of our results with $SU(3)_F$ and those in the literature for the up-down asymmetries of  
			the singly Cabibbo-suppressed charmed baryon decays,
			where  UVK, SV2 and  CKX are from
			Refs.~\cite{Uppal:1994pt}, \cite{Sharma:1996sc} and \cite{Cheng:sup}, respectively.
		}\label{Table_alpha_sup}
		\begin{tabular}[t]{lcccc}
			\hline
			channel & our result & UVK$^{(\prime)}$&SV2$^{(\prime)}$&CKX  \\
			\hline
			$ \Lambda_{c}^{+}  \to  p \pi^{0} $&~$ -0.05 \pm 0.72$~             &$0.82~(0.85)$& $0.05~(0.05)$&~$-0.95$~\\
			$ \Lambda_{c}^{+}  \to  p \eta $&$ -0.94^{+0.26}_{-0.06}$     &~$-1.00~(-0.79)$~&~$-0.74~(-0.45)$~&~$-0.56$~\\
			$ \Lambda_{c}^{+}  \to  p \eta' $&$ 0.91 ^{+ 0.09 }_{- 0.21 }$ &$0.87~(0.87)$&$-0.97~(-0.99)$&\\
			$ \Lambda_{c}^{+}  \to  n \pi^{+} $&$ 0.12 \pm 0.19$                    &$-0.13~(0.68)$&$0.05~(0.05)$&$-0.90$\\
			$ \Lambda_{c}^{+}  \to  \Lambda^{0} K^{+} $&$ 0.32 \pm 0.32$              &$-0.99~(-0.99)$&$-0.54~(0.97)$&$-0.96$\\
			$ \Lambda_{c}^{+}  \to  \Sigma^{0} K^{+} $&$ -1.00 ^{+ 0.06 }_{- 0.00 }$ &$-0.80~(-0.80)$&$0.68~(-0.98)$&$-0.73$ \\
			$ \Lambda_{c}^{+}  \to  \Sigma^{+} K^0 $&$ -1.00 ^{+ 0.06 }_{- 0.00 }$    &$-0.80~(-0.80)$&$0.68~(-0.98)$&$-0.74$\\
			\hline
		\end{tabular}
		
	\end{center}
\end{table}
In the tables, the data are taken from the experimental values in Ref.~\cite{pdg},
KK and Iva correspond to the calculations with the covariant quark models by
Korner and Kramer (KK)~\cite{Korner92} and Ivanov $el~al.$ (Iva)~\cite{xialpha1}, 
 XK, CT and Zen are  based on the pole models by
 Xu and Kamal (XK)~\cite{Xu92}, Cheng and Tseng (CT)~\cite{Cheng:allow} and
Zenczykowski (Zen)~\cite{xialpha2}, 
 SV1, CT$^{\prime}$, UVK$^{(\prime)}$  and  CKX 
are related to the considerations of current algebra by  Sharma and Verma (SV1)~\cite{Verma98},  Cheng and Tseng (CT)~\cite{Cheng:allow},
Uppal, Verma and Khanna (UVK) without (with) the baryon wave function scale 
variation~\cite{Uppal:1994pt} 
 and Cheng, Kang and Xu (CKX)~\cite{Cheng:sup},
and SV2$^{(\prime)}$  represent the results with $SU(3)_F$ by Sharma and Verma  with two different signs of $B(\Lambda_c^+\to \Xi^0 K^+)$~\cite{Sharma:1996sc},  respectively.
As seen in Table~\ref{Table_alpha}, our results  of the up-down asymmetries are quite different from those in the 
literature~\cite{Korner92,Xu92,xialpha1,xialpha2,Cheng:allow,Cheng:general,Uppal:1994pt,Verma98,Sharma:1996sc}.
In particular, it is interesting to see that we predict that
\begin{equation}\label{alpha}
  \alpha(\Lambda_c^+\to \Xi^0 K^+)_{SU(3)} = 0.94^{+0.06}_{-0.11}\,
\end{equation}
which is consistent with the current experimental data of $0.77\pm0.78$ in Eq.~(\ref{expal})~\cite{Ablikim:2018bir}, 
but  different from all theoretical predictions in the literature.
For example, it has been suggested that this asymmetry is approximately zero in 
dynamical models~\cite{Korner92,Xu92,xialpha1,xialpha2,Cheng:allow,Cheng:general,Uppal:1994pt,Verma98}, while
the authors in  Ref.~\cite{Sharma:1996sc} have  also taken it to be zero as a data input when  the $SU(3)_F$ symmetry is imposed.
In our fit, the value in Eq.~(\ref{expal}) has not been included as an input in order to see its value  based on the $SU(3)_F$ approach.
Since the error of our predicted result in Eq.~(\ref{alpha}) is small, we are confident that $\alpha(\Lambda_c^+\to \Xi^0 K^+)$ should be 
much lager than zero and close to one. 
Moreover,
 our result of
$\alpha(\Lambda_c^+\to \Lambda_0K^+)_{SU(3)}=0.32\pm 0.32$ 
is different from 
the CKX one of  $\alpha(\Lambda_c^+\to \Lambda_0K^+)_{CKX}=-0.96$ in Ref.~\cite{Cheng:sup}. 
The reason for the difference is 
due to the  signs in the parity violated amplitudes of
  $A_{(\Lambda_c^+\to \Lambda_0K^+)_{SU(3)}}=(1.5\pm 0.1)\times10^{-2}G_F\text{GeV}^2$ in our calculation
  and  $A_{(\Lambda_c^+\to \Lambda_0K^+)_{CKX}}=-1.57\times10^{-2}G_F\text{GeV}^2$ in Ref.~\cite{Cheng:sup}.  
  To clarify these issues,  further precision measurements on these asymmetries are highly recommended.
 
 In addition, due to the vanishing contributions to the decays from the $a_4$, $a_5$, $a_7$ and $a'_0$ terms of ${\cal O}(\overline{15})$, we
 get
 \begin{eqnarray}\label{Result1}
   A{(\Lambda_c^+ \to\Sigma^0 K^+)} &=& A{(\Lambda_c^+\to \Sigma^+ K^0_S, \Sigma^+K^0_L)}=\sqrt{2}(a_1 -a_3 )s_c\,,\nonumber\\
   B{(\Lambda_c^+ \to\Sigma^0 K^+)} &=& B{(\Lambda_c^+\to \Sigma^+ K^0_S, \Sigma^+K^0_L)}=\sqrt{2}(b_1 -b_3 )s_c\,,
   \end{eqnarray}
leading to the fitted values of 
\begin{eqnarray}\label{Result1a}
&&{\cal B}(\Lambda_c^+\to \Sigma^0 K^+, \Sigma^+ K^0_S, \Sigma^+K^0_L)=(5.4\pm 0.7)\times 10^{-4}\,,
\nonumber\\
&&\alpha(\Lambda_c^+\to \Sigma^0 K^+, \Sigma^+ K^0_S, \Sigma^+K^0_L)= -1.00 ^{+ 0.06 }_{- 0.00 }\,,
\end{eqnarray}
as given in Table~\ref{Table3}.
  Note  that the decay branching ratio of $\Lambda_c^+ \to\Sigma^0 K^+$ has been measured to be $(5.2\pm 0.8)\times 10^{-4}$~\cite{pdg},
  which agrees with with that in Eq.~(\ref{Result1a}).
  Future measurements on $\Lambda_c^+\to \Sigma^+ K^0_S$ and  $\Lambda_c^+\to \Sigma^+K^0_L$ are important as they can tell us if 
  Eqs.~(\ref{Result1}) and (\ref{Result1a}),
  which can also be derived through the isospin symmetry, are right or wrong.

We now concentrate on the decay processes of ${\bf B}_c \to {\bf B}_n K_L^0$ and ${\bf B}_c \to {\bf B}_n K_S^0$, which involve both  Cabibbo-allowed and   doubly suppressed transitions,
as shown in Table~\ref{Table5}. 
\begin{table}
\begin{center}
  \centering
  \caption{Irreducible amplitudes, decay branching ratios and up-down and  $K_S^0-K_L^0$ asymmetries of ${\bf B}_c \to {\bf B}_n K_L^0/K_S^0$ with both  Cabibbo-allowed and   doubly Cabibble-suppressed transitions, where the $B$  amplitudes can be obtained directly by substituting $a_i$ with $b_i$.}\label{Table5}
\begin{tabular}[t]{ccccc}
\hline
channel & Irreducible amplitude for A&$10^3{\cal B}_{SU(3)_F}$&$\alpha_{SU(3)_F}$&${\bf R}_{K^0_{S,L}}$\\
\hline
$ \Lambda_{c}^{+}  \to  p K_S^0 $ & $  \sqrt{2}\left( (a_{1} - \frac{ a_{6}}{2})+ (a_{3}  - \frac{ a_{6}}{2}) s_c^{2}  \right) $ &$ 15.7 \pm 0.8 $&$-0.89^{+0.26}_{-0.11}$&$0.009\pm 0.011$\\
$ \Lambda_{c}^{+}  \to  p K_L^0 $ & $ -\sqrt{2}\left((a_{1} - \frac{ a_{6}}{2})- (a_{3} - \frac{ a_{6}}{2})s_c^{2}  \right) $ &$ 15.5 \pm 0.8 $&$-0.92^{+0.21}_{-0.08}$&\\
\hline
$ \Xi_{c}^{+}  \to  \Sigma^{+} K_S^0 $ & $ -\sqrt{2}\left( (a_{3}- \frac{ a_{6}}{2})+ ( a_{1}  - \frac{ a_{6}}{2} )s_c^{2} \right) $ &$ 4.9^{+ 5.9}_{-4.2} $&$0.89^{+0.11}_{-0.46}$&$0.118\pm 0.078$\\
$ \Xi_{c}^{+}  \to  \Sigma^{+} K_L^0 $ & $ \sqrt{2}\left(  (a_{3}- \frac{ a_{6}}{2}) -  (a_{1}   - \frac{ a_{6}}{2} )s_c^{2} \right) $ &$ 3.9^{+5.1}_{-3.5} $&$1.00^{+0.00}_{-0.18}$&\\
\hline
$ \Xi_{c}^{0}  \to  \Sigma^{0} K_S^0 $ & $ (a_{2} + a_{3} - \frac{a_{6}}{2})+ (a_{1} - \frac{a_{6}}{2})s_c^{2}  $ &$ 0.5\pm 0.4$&$-0.34 ^{+ 0.95}_{-0.66}$&$ 0.170 \pm 0.146$\\
$ \Xi_{c}^{0}  \to  \Sigma^{0} K_L^0 $ & $  -( a_{2} + a_{3}- \frac{a_{6}}{2})+(a_{1} - \frac{a_{6}}{2})s_c^{2} $ &$ 0.3^{+0.5}_{-0.3}$&$0.28\pm0.71$&\\
\hline
$ \Xi_{c}^{0}  \to  \Lambda^{0} K_S^0 $ & $\frac{1}{\sqrt{3}}( ( 2 a_{1} - a_{2} - a_{3}- \frac{ a_{6}}{2})$&$ 5.0 \pm 0.3 $&$-0.70\pm0.28$&$-0.043\pm 0.003$\\
& $ -(a_{1}- 2a_{2}  - 2 a_{3} +\frac{ a_{6}}{2})s_c^{2}) $ &&&\\
$ \Xi_{c}^{0}  \to  \Lambda^{0} K_L^0 $ & $ -\frac{1}{\sqrt{3}}( (2 a_{1}  -  a_{2} - a_{3}- \frac{ a_{6}}{2})$&$ 5.5\pm 0.3 $&$-0.66\pm0.28$&\\
& $+(a_{1}- 2a_{2}  - 2 a_{3} +\frac{ a_{6}}{2})s_c^{2}) $ &&&\\
\hline
\end{tabular}
\end{center}
\end{table}
 If we ignore the later contributions associated with $\sin^2\theta_c$, ${\cal B}({\bf B}_c \to {\bf B}_n K_S^0)
={\cal B}({\bf B}_c \to {\bf B}_n K_L^0)$. Clearly, the $K_S^0-K_L^0$ asymmetry depends on the doubly Cabibbo-suppressed parts of the decays.
As shown in Table~\ref{Table5}, the central values for the first three asymmetries are predicted  to be around $10\%$ or more,
which are consistent with those in Ref.~\cite{Wang:2017gxe}.
For $\Xi_c^0 \to \Sigma^0 K_S^0/K_L^0$,
 the up-down asymmetry  
 of  ${\bf R}_{K^0_{S,L}}(\Xi_c^0 \to \Sigma^0) $ 
has different sign, indicating that the effect of 
the doublyCabibbo-suppressed transition is not ignorable in these decay processes. 
Explicitly, we find out that  ${\cal B}(\Xi_c^0 \to \Lambda^0 K_L^0)$ can be a little larger than ${\cal B}(\Xi_c^0 \to \Lambda^0 K_S^0)$, 
in which  the $K_S^0-K_L^0$ asymmetry is predicted to be $-(4.3\pm0.3)\%$ with a tiny uncertainty, which agrees well with $-(3.7\pm0.4)\%$
in Ref.~\cite{Wang:2017gxe}.

\section{Conclusions}

We have studied the two-body decays of ${\bf B}_c \to {\bf B}_n M$ with the $SU(3)_F$ flavor symmetry based on the spin-dependent $s$ and
$p$-wave amplitudes
of $A$ and $B$, respectively. These 
amplitudes, which have been decomposed in terms of the  $SU(3)_F$ parameters $a_i^{(\prime)}$ and $b_i^{(\prime)}$, 
allow us to examine the longitudinal polarization of $P_{{\bf B}_n}$, which is related to the up-down asymmetry of
$\alpha$.
We have obtained a good $\chi$ fit for the ten $SU(3)_F$ parameters in Eq.~(\ref{fittingresults}) from
 the all possible contributions of ${\cal O}(6)$ and ${\cal O}(\overline{15})$ 
 with 16 data points in Table~\ref{Table1} in the  $SU(3)_F$ approach, 
 in which all
experimental data for the decay branching ratios and  up-down asymmetries can be explained.
Consequently, we have systematically predicted all decay branching ratios and up-down asymmetries of
the Cabibbo-allowed, singly Cabibbo-suppressed and doubly Cabibbo-suppressed charmed baryon decays.
In particular, our results of  ${\cal B}( \Xi_{c}^{0}  \to  \Xi^{-} \pi^{+})=(2.21\pm0.14)\times 10^{-2} $ and
 $\alpha( \Xi_{c}^{0}  \to  \Xi^{-} \pi^{+})=-0.98^{+0.07}_{-0.02}$ are consistent with the  data of
$(1.80\pm0.52)\times 10^{-2} $~\cite{Exp:absXic0}
 and $-0.6\pm0.4$~\cite{pdg}, respectively. We have also found that
 ${\cal B}(\Lambda_c^+\to p \pi^0)=(1.2\pm1.2)\times 10^{-4}$, which is consistent  with 
   the current experimental upper limit of $2.7\times 10^{-4}$~\cite{pdg}. 
   In addition, we have gotten that
   ${\cal B}(\Lambda_c^+\to \Sigma^0 K^+, \Sigma^+ K^0_S, \Sigma^+K^0_L)=(5.4\pm 0.7)\times 10^{-4}$
   and $\alpha(\Lambda_c^+\to \Sigma^0 K^+, \Sigma^+ K^0_S, \Sigma^+K^0_L)= -1.00 ^{+ 0.06 }_{- 0.00 }$,
   which are also guaranteed by the isospin symmetry.
   
  We have shown in Table~\ref{Table_alpha} that  our predictions of the up-down asymmetries are quite different from the theoretical values in the 
literature for most of the decay modes. 
 In particular, we have found that $ \alpha(\Lambda_c^+\to \Xi^0 K^+)_{SU(3)} = 0.94^{+0.06}_{-0.11}$ in Eq.~(\ref{alpha}),
which is consistent with the current experimental data of $0.77\pm0.78$ in Eq.~(\ref{expal})~\cite{Ablikim:2018bir}, 
but much larger than zero predicted in the literature. A future precision measurement on this asymmetry is clearly very important
as our prediction based on $SU(3)_F$ is close to one with a small uncertainty, which can be viewed as a benchmark for the $SU(3)_F$ approach.

We have also explored the $K_S^0-K_L^0$asymmetries  in the decays of ${\bf B}_c \to {\bf B}_n K_L^0/K_S^0$ with  both  Cabibbo-allowed and   doubly  Cabibbo-suppressed transitions. The asymmetries depend strongly on the contributions from the doubly  Cabibbo-suppressed contributions.
Clearly, the measurements of these asymmetries are good tests for the doubly  Cabibbo-suppressed transitions.

In conclusion, we give a systematic consideration of the up-down asymmetries in the two-body charmed baryon decays of ${\bf B}_c \to {\bf B}_n M$  as well as the $K_S^0-K_L^0$asymmetries  in the decays of ${\bf B}_c \to {\bf B}_n K_L^0/K_S^0$ in the $SU(3)_F$ approach.
 Some of our predictions based on $SU(3)_F$ are different from those in the dynamical models, can be tested  by the  experiments at BESIII and Belle.

\appendix
\section{Irreducible Amplitudes}
In this Appendix, 
we provide the  irreducible amplitudes $A_{{\bf B}_c\to{\bf B}_nM}$ from Eq.~(\ref{abamp})
based on the flavor $SU(3)_F$ symmetry, while those of $B_{{\bf B}_c\to{\bf B}_nM}$ can be obtained by substituting $b_i$ with $a_i$
in $A_{{\bf B}_c\to{\bf B}_nM}$  . 
Note that in  the limits of $\eta=\eta_8$ and $\eta'=\eta_0$,  one has that $s_\phi=\sqrt{2}c_\phi$, resulting in
 the $\eta'=\eta_0$ modes  only contain $\tilde{a}$.
 In Tables~\ref{Table6}, \ref{Table7} and \ref{Table8},
 we show the Cabibbo-allowed, singly Cabibbo-suppressed and doubly Cabibbo-suppressed amplitudes of $A_{{\bf B}_c \to {\bf B}_n M}$, respectively.
 Here, we have only considered the factorizable amplitudes from ${\cal O}(\overline{15})$, so that the terms associated with
 $a_{0,4,5,7}$ and $b_{0,4,5,7}$ are set to be zero.
 
\begin{table}[t]
	\begin{center}
		\centering
		\caption{ Cabibbo-allowed amplitudes for $A_{{\bf B}_c \to {\bf B}_n M}$ }\label{Table6}
		\begin{tabular}[t]{lc}
			\hline
			Channel & $A$ \\
			\hline
			$ \Lambda_{c}^{+}  \to  \Lambda^{0} \pi^{+} $ & $ \frac{\sqrt{6}}{3}(- a_{1} - a_{2} - a_{3} - a_{6})$ \\
			\hspace{0.62cm}$\to  p \bar{K}^{0} $ & $ - 2 a_{1} + a_{6} $ \\
			\hspace{0.62cm}$   \to  \Sigma^{0} \pi^{+} $ & $ \sqrt{2}(- a_{1} + a_{2} + a_{3}) $ \\
			\hspace{0.62cm}$   \to  \Sigma^{+} \pi^{0} $ & $ \sqrt{2}( a_{1} - a_{2} - a_{3})$ \\
			\hspace{0.62cm}$  \to  \Sigma^{+} \eta $ & $ \frac{\sqrt{2}}{3}c_{\phi}(- a_{1} - a_{2} +a_{3} -6 \tilde{a})+\frac{2}{3}s_{\phi}(-  a_{1} -  a_{2} +  a_{3} + 3 \tilde{a})$ \\
			\hspace{0.62cm}$   \to  \Sigma^{+} \eta' $ & $\frac{2}{3} c_{\phi}( a_{1} +  a_{2} -  a_{3} - 3 \tilde{a})+ \frac{\sqrt{2}}{3}s_{\phi}(- a_{1} - a_{2} + a_{3} - 3\tilde{a})$ \\
			\hspace{0.62cm}$   \to  \Xi^{0} K^{+} $ & $ - 2 a_{2} $ \\
			\hline
			$ \Xi_{c}^{+}  \to    \Sigma^{+} \bar{K}^{0} $ & $ 2 a_{3} - a_{6} $ \\
			\hspace{0.62cm}$ \to  \Xi^{0} \pi^{+} $ & $  -2 a_{3} - a_{6} $ \\
			\hline
			$ \Xi_{c}^{0}\to  \Lambda^{0} \bar{K}^{0} $ & $ \frac{\sqrt{6}}{3}(- 2 a_{1} + a_{2} + a_{3} + \frac{a_{6}}{2}) $ \\
			\hspace{0.52cm}$  \to \Sigma^{0} \bar{K}^{0} $ & $ \sqrt{2}(- a_{2} - a_{3} + \frac{a_{6}}{2}) $ \\
			\hspace{0.52cm}$  \to  \Sigma^{+} K^{-} $ & $ 2 a_{2} $ \\
			\hspace{0.52cm}$ \to  \Xi^{0} \pi^{0} $ & $  \sqrt{2}(- a_{1} + a_{3})$ \\
			\hspace{0.52cm}$   \to  \Xi^{0} \eta $ & $  \frac{\sqrt{2}}{3}c_{\phi}(a_{1} - 2 a_{2} - a_{3}+ 6 \tilde{a})+\frac{2}{3}s_{\phi}( a_{1} - 2 a_{2} -  a_{3} - 3 \tilde{a}) $ \\
			\hspace{0.52cm}$  \to  \Xi^{0} \eta' $ & $ \frac{2}{3}c_{\phi}(- a_{1} + 2 a_{2} + a_{3} + 3 \tilde{a})+\frac{\sqrt{2}}{3}s_{\phi}(a_{1} - 2 a_{2}- a_{3} + 6 \tilde{a})$ \\
			\hspace{0.52cm}$   \to  \Xi^{-} \pi^{+} $ & $ 2 a_{1} + a_{6} $ \\
			\hline
		\end{tabular}
	\end{center}
\end{table}
\begin{table}
	\renewcommand\arraystretch{0.9}
	\begin{center}
		\centering
		\caption{Singly Cabibbo-suppressed amplitudes for $A_{{\bf B}_c \to {\bf B}_n M}$ }\label{Table7}
		\begin{tabular}[t]{lc}
			\hline
			Channel & $\sin^{-1}\theta_c A$\\
						\hline
			$ \Lambda_{c}^{+}  \to  \Lambda^{0} K^{+} $ & $\frac{\sqrt{6}}{3}(a_{1} - 2 a_{2} +a_{3} +a_{6}) $ \\
			\hspace{0.62cm}$  \to  p \pi^{0} $ & $ \sqrt{2}(a_{2} + a_{3} - \frac{a_{6}}{2}) $ \\
\hspace{0.62cm}$ \to  p \eta $ & $ \sqrt{2}c_{\phi}( - \frac{2 a_{1}}{3} + \frac{a_{2}}{3} + \frac{a_{3}}{3} + \frac{a_{6}}{2}+2 \tilde{a})+\frac{1}{3}s_{\phi}( - 4 a_{1} + 2 a_{2}+ 2 a_{3}+ 3a_{6}- 6 \tilde{a}) $ \\
\hspace{0.62cm}$ \to  p \eta' $ & $ \frac{1}{3}c_{\phi}( 4 a_{1} - 2 a_{2} - 2 a_{3} - 3a_{6}+6 \tilde{a})+\sqrt{2}s_{\phi}( - \frac{2 a_{1}}{3} + \frac{a_{2}}{3} + \frac{a_{3}}{3} + \frac{a_{6}}{2}+2 \tilde{a}) $ \\

			\hspace{0.62cm}$  \to  n \pi^{+} $ & $ 2 a_{2} + 2 a_{3} + a_{6} $ \\
			 \hspace{0.62cm}$ \to  \Sigma^{0} K^{+} $ & $ \sqrt{2}(a_{1} - a_{3}) $ \\
			\hspace{0.62cm}$  \to  \Sigma^{+} K_S $ & $ \sqrt{2}(a_{1} - a_{3}) $ \\
			\hspace{0.62cm}$ \to  \Sigma^{+} K_L $ & $ \sqrt{2}(a_{1} - a_{3}) $ \\
			\hline
			$ \Xi_{c}^{+}  \to  \Lambda^{0} \pi^{+} $ & $\frac{\sqrt{6}}{3}(a_{1} + a_{2} -2 a_{3} - \frac{a_{6}}{2})$ \\
\hspace{0.62cm}$ \to  p K_S $ & $ \sqrt{2}(- a_{1} + a_{3}) $ \\
\hspace{0.62cm}$ \to  p K_L $ & $ \sqrt{2}(a_{1} - a_{3}) $ \\
\hspace{0.62cm}$ \to  \Sigma^{0} \pi^{+} $ & $ \sqrt{2}(a_{1} - a_{2} + \frac{a_{6}}{2}) $ \\
\hspace{0.62cm}$ \to  \Sigma^{+} \pi^{0} $ & $ \sqrt{2}(- a_{1} + a_{2} + \frac{a_{6}}{2}) $ \\
\hspace{0.62cm}$ \to  \Sigma^{+} \eta $ & $ \sqrt{2}c_{\phi}( \frac{a_{1}}{3} + \frac{a_{2}}{3} + \frac{2 a_{3}}{3} - \frac{a_{6}}{2}+2 \tilde{a})+\frac{1}{3}s_{\phi}( 2 a_{1} + 2 a_{2} + 4 a_{3} -3 a_{6}- 6 \tilde{a}) $ \\
\hspace{0.62cm}$\to  \Sigma^{+} \eta' $ & $ \frac{1}{3}c_{\phi}( - 2 a_{1} - 2 a_{2} - 4 a_{3} + 3a_{6}+6 \tilde{a})+\sqrt{2}s_{\phi}( \frac{a_{1}}{3} + \frac{a_{2}}{3} + \frac{2 a_{3}}{3} - \frac{a_{6}}{2}+2 \tilde{a}) $ \\
			\hspace{0.62cm}$  \to  \Xi^{0} K^{+} $ & $ 2 a_{2} + 2 a_{3} + a_{6} $ \\
			\hline
$ \Xi_{c}^{0}  \to  \Lambda^{0} \pi^{0} $ & $ \frac{\sqrt{3}}{3}(- a_{1}- a_{2} + 2 a_{3} - \frac{a_{6}}{2}) $ \\
\hspace{0.52cm}$\to  \Lambda^{0} \eta $ & $\frac{\sqrt{3}}{3} c_{\phi}( - a_{1} - a_{2} + \frac{a_{6}}{2}+6 \tilde{a})+\frac{\sqrt{6}}{3}s_{\phi}( - a_{1} - a_{2} + \frac{a_{6}}{2}- 3\tilde{a}) $ \\
\hspace{0.52cm}$  \to  \Lambda^{0} \eta' $ & $ \frac{\sqrt{6}}{3}c_{\phi}(a_{1} + a_{2} - \frac{a_{6}}{2}+3\tilde{a})+\frac{\sqrt{3}}{3}s_{\phi}( - a_{1} - a_{2} + \frac{a_{6}}{2}+6 \tilde{a}) $ \\
\hspace{0.52cm}$\to  p K^{-} $ & $ - 2 a_{2} $ \\
\hspace{0.52cm}$ \to  n K_S $ & $ \sqrt{2}(- a_{1} + a_{2} + a_{3}) $ \\
\hspace{0.52cm}$\to  n K_L $ & $ \sqrt{2}(a_{1} - a_{2} - a_{3}) $ \\
\hspace{0.52cm}$  \to  \Sigma^{0} \pi^{0} $ & $ a_{1} + a_{2} - \frac{a_{6}}{2} $ \\
\hspace{0.52cm}$ \to  \Sigma^{0} \eta $ & $ c_{\phi}( - \frac{a_{1}}{3} - \frac{a_{2}}{3} - \frac{2 a_{3}}{3} + \frac{a_{6}}{2}- 2 \tilde{a})+\sqrt{2}s_{\phi}(- \frac{a_{1}}{3} - \frac{a_{2}}{3} - \frac{2 a_{3}}{3} + \frac{a_{6}}{2}+\tilde{a} ) $ \\
\hspace{0.52cm}$ \to  \Sigma^{0} \eta' $ & $ \sqrt{2}c_{\phi}( \frac{a_{1}}{3} + \frac{a_{2}}{3} + \frac{2 a_{3}}{3} - \frac{a_{6}}{2}- \tilde{a})+s_{\phi}(- \frac{a_{1}}{3} - \frac{a_{2}}{3} - \frac{2 a_{3}}{3} + \frac{a_{6}}{2}- 2 \tilde{a} ) $ \\
\hspace{0.52cm}$ \to  \Sigma^{+} \pi^{-} $ & $ 2 a_{2} $ \\
\hspace{0.52cm}$\to  \Sigma^{-} \pi^{+} $ & $ 2 a_{1} + a_{6} $ \\
\hspace{0.52cm}$ \to  \Xi^{0} K_S $ & $ \sqrt{2}(- a_{1} + a_{2} + a_{3}) $ \\
\hspace{0.52cm}$ \to  \Xi^{0} K_L $ & $ \sqrt{2}(- a_{1} + a_{2} + a_{3}) $ \\
\hspace{0.52cm}$  \to  \Xi^{-} K^{+} $ & $ - 2 a_{1} - a_{6} $ \\
			\hline
		\end{tabular}
	\end{center}
\end{table}
\begin{table}
	\renewcommand\arraystretch{1}
	\begin{center}
		\centering
		\caption{Doubly Cabibbo-suppressed amplitudes for $A_{{\bf B}_c \to {\bf B}_n M}$ }\label{Table8}
		\begin{tabular}[t]{lc}
			\hline
			Channel & $\sin^{-2}\theta_c A$\\
			\hline
			$ \Lambda_{c}^{+}  \to  p K^{0} $ & $ 2 a_{3} - a_{6} $ \\
			\hspace{0.62cm}$ \to  n K^{+} $ & $ - 2 a_{3} - a_{6} $ \\
			\hline
			$ \Xi_{c}^{+}  \to  \Lambda^{0} K^{+} $ & $ \frac{\sqrt{6}}{3}(- a_{1} + 2 a_{2} + 2 a_{3}+ \frac{a_{6}}{2}) $ \\
			\hspace{0.62cm}$ \to  p \pi^{0} $ & $ -\sqrt{2}a_{2} $ \\
			\hspace{0.62cm}$  \to  p \eta $ & $ \frac{\sqrt{2}}{3}c_{\phi}(  2 a_{1}- a_{2} - 2 a_{3}- 6 \tilde{a})+\frac{2}{3}s_{\phi}( 2 a_{1} -  a_{2} - 2 a_{3}+3 \tilde{a}) $ \\
			\hspace{0.62cm}$  \to  p \eta' $ & $ \frac{2}{3}c_{\phi}( - 2 a_{1} +  a_{2} + 2 a_{3}- 3\tilde{a})+\frac{\sqrt{2}}{3}s_{\phi}( 2 a_{1} -a_{2} - 2 a_{3}- 6 \tilde{a} ) $ \\
			\hspace{0.62cm}$\to  n \pi^{+} $ & $ - 2 a_{2} $ \\
			\hspace{0.62cm}$\to  \Sigma^{0} K^{+} $ & $ \sqrt{2}(- a_{1} - \frac{a_{6}}{2}) $ \\
			\hspace{0.62cm}$\to  \Sigma^{+} K^{0} $ & $ - 2 a_{1} + a_{6} $ \\
			\hline
			$ \Xi_{c}^{0}  \to  \Lambda^{0} K^{0} $ & $ \frac{\sqrt{6}}{3}(- a_{1} + 2 a_{2} + 2 a_{3} - \frac{a_{6}}{2}) $ \\
			\hspace{0.52cm}$  \to  p \pi^{-} $ & $ - 2 a_{2} $ \\
			\hspace{0.52cm}$ \to  n \pi^{0} $ & $ \sqrt{2}a_{2} $ \\
			\hspace{0.52cm}$ \to  n \eta $ & $ \frac{\sqrt{2}}{3}c_{\phi}(- 6 \tilde{a} + 2 a_{1} - a_{2} - 2 a_{3})+\frac{2}{3}s_{\phi}(3 \tilde{a} + 2 a_{1} -  a_{2} -2 a_{3}) $ \\
			\hspace{0.52cm}$\to  n \eta' $ & $ \frac{2}{3}c_{\phi}(- 3 \tilde{a} - 2 a_{1} +  a_{2} + 2 a_{3})+\frac{\sqrt{2}}{3}s_{\phi}(- 6 \tilde{a} +2 a_{1} - a_{2} - 2 a_{3}) $ \\
				\hspace{0.52cm}$\to  \Sigma^{0} K^{0} $ & $ \sqrt{2}(a_{1} - \frac{a_{6}}{2}) $ \\
			\hspace{0.52cm}$\to  \Sigma^{-} K^{+} $ & $ - 2 a_{1} - a_{6} $ \\
			\hline
		\end{tabular}
	\end{center}
\end{table}

\section{Up-down and Longitudinal Polarization Asymmetries }

From Eq.~(\ref{Angular}), the up-down is defined by
\begin{eqnarray}
\alpha &=& 
{d\Gamma (\vec{P}_{{\bf B}_n}\cdot \hat{p}_{{\bf B}_n}=+1)-d\Gamma (\vec{P}_{{\bf B}_n}\cdot \hat{p}_{{\bf B}_n}=-1) \over
d\Gamma (\vec{P}_{{\bf B}_n}\cdot \hat{p}_{{\bf B}_n}=+1)+d\Gamma (\vec{P}_{{\bf B}_n}\cdot \hat{p}_{{\bf B}_n}=-1) }\,,
\end{eqnarray}
which is equal to the longitudinal polarization asymmetry, $i.e.$ $P_{{\bf B}_n}=\alpha$.

\section*{ACKNOWLEDGMENTS}
This work was supported in part by National Center for Theoretical Sciences and
MoST (MoST-104-2112-M-007-003-MY3 and MoST-107-2119-M-007-013-MY3).

\end{document}